\documentclass[twocolumn,showpacs,preprintnumbers,amsmath,amssymb]{revtex4-1}

\usepackage{graphicx}
\usepackage{dcolumn}
\usepackage{bm}

\begin{document}

\title{Evolution of Paramagnetic Quasiparticle Excitations Emerged in the High-Field Superconducting Phase  of CeCoIn$_5$}

\author{K.~Kumagai$^{1}$}
\author{H.~Shishido$^{2}$}
\author{T.~Shibauchi$^{2}$}
\author{Y.~Matsuda$^{2}$}

\affiliation{$^{1}$Department of Physics, Hokkaido University, Sapporo 060-0810, Japan}

\affiliation{$^{2}$Department of Physics, Kyoto University, Kyoto 606-8502, Japan}


\begin{abstract}

We present $^{115}$In NMR measurements in a novel thermodynamic phase of CeCoIn$_5$ in high magnetic field, where exotic superconductivity coexists with the incommensurate spin-density wave order. We show that the NMR spectra in this phase provide direct evidence for the emergence of the spatially distributed normal quasiparticle regions.  The quantitative analysis for the field evolution of the paramagnetic  magnetization and newly-emerged low-energy quasiparticle density of states is consistent with the nodal plane formation, which is characterized by an order parameter in the Fulde-Ferrell-Larkin-Ovchinnikov (FFLO) state.  The NMR spectra also suggest that the spatially uniform spin-density wave is induced in the FFLO phase. 

\end{abstract}

\maketitle

\begin{figure}[tb]
\includegraphics[width=0.97\linewidth]{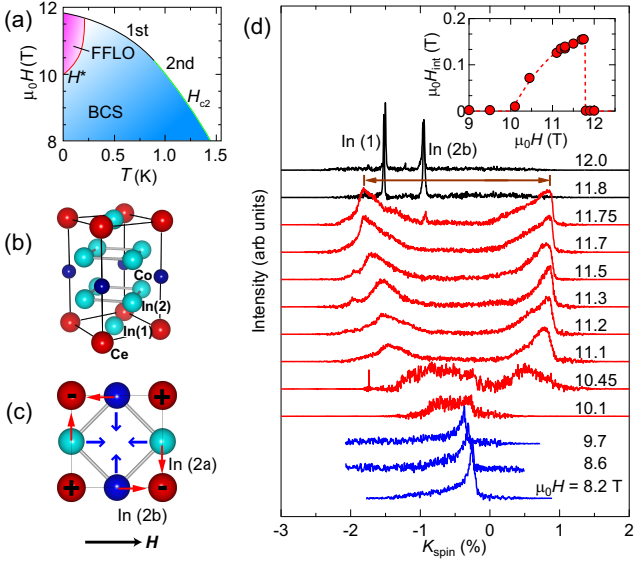}
\caption{
(Color online)
(a) $H$-$T$ phase diagram of CeCoIn$_5$ for {\boldmath $H$}\,$\parallel ab$. (b) Crystal structure of CeCoIn$_5$. (c) A top view of the In(2) plane, and directions of the largest principal axis of electric field gradient (blue arrows) and of hyperfine fields (red arrows) on the In(2a) (cyan circles) and the In(2b) (blue circles) sites transferred from the magnetic Ce moments parallel to the $c$ axis. (d) Field evolution of the NMR spectra at the In(2b) site as a function of $K_{\rm spin}$ at $T=0.05$\,K in the normal (black), HL (red), and BCS (blue) states. $K_{\rm spin}=K-K_{\rm orb}$ is estimated by using $K_{\rm orb}=2.1\%$. Inset: Field dependence of the internal field $H_{\rm int}=\delta f /2\gamma_N$ obtained by the difference of the resonance frequency $\delta f$ of the two peaks (double-headed arrow in the main panel).   
}
\label{fig:SDW}
\end{figure}

The interplay between magnetism and unconventional superconductivity with a nontrivial Cooper pairing has been a topic of recent intense study. The quasi-two dimensional heavy fermion superconductor CeCoIn$_5$ \cite{Pet01} continues to excite great interest, because it shows a number of fascinating superconducting properties \cite{Kas05,Iza01,Tayama,Bia02,Whi10,Rad03,Bia03}. Its superconductivity at high fields is destroyed by Pauli paramagnetic effect, as evidenced by the the first-order phase transition at the upper critical field $H_{c2}$ \cite{Iza01,Tayama,Bia02} and anomalous flux line lattice form factor \cite{Whi10}.  What is striking is that CeCoIn$_5$ exhibits a new thermodynamic phase transition at ($T^*,H^*$) just below $H_{c2}$ (Fig.\:\ref{fig:SDW}(a)) for both {\boldmath $H$}\,$\parallel ab$  and {\boldmath $H$}\,$\parallel c$ \cite{Rad03,Bia03}.   Closely related to the Pauli limited superconductivity,  this high-field and low-temperature (HL) phase has been attributed to the Fulde-Ferrell-Larkin-Ovchinnikov (FFLO) state \cite{Fulde,Larkin}, in which the pair-breaking arising from the Pauli effect is reduced by forming a new pairing state  ({\boldmath $k$}$\uparrow$,{\boldmath $-k+q$}$\downarrow$)  with nonzero {\boldmath $q$} between the Zeeman splitted parts of the Fermi surface.  One of the most fascinating aspects of the FFLO state is that Cooper pairs with finite center-of-mass momenta $\hbar${\boldmath $q$} develop an oscillating superconducting order parameter in real space such as  $\Delta$({\boldmath $r$}) $\propto \sin$({\boldmath $q$}$\cdot${\boldmath $r$}) and, as a result,  nodal planes appear periodically perpendicular to the applied field \cite{Mat07}.  

The presence of the FFLO state in CeCoIn$_5$ has been supported by several experiments, including impurity \cite{Tok08}, pressure \cite{Mic06}, and ultrasound studies \cite{Wat04,Ike07}.  However, recent NMR \cite{You07,Cur10,Kou10} and neutron \cite{Ken08,Ken10} data in parallel field demonstrated a long-range static magnetic order in the HL phase. The magnetic moment at Ce atoms is given by {\boldmath $\mu$}({\boldmath $r$}) $=$ {\boldmath $\mu$}$_0\cos$({\boldmath $Q$}$_{S}\cdot${\boldmath $r$}) with {\boldmath $Q$}$_{S}= 2\pi\left(\frac{\delta}{a},\frac{\delta}{a},\frac{0.5}{c}\right)$ $(\delta\sim0.45)$ \cite{Ken08}, which is directed to the $c$ axis, {\boldmath $\mu$}$_0\parallel c$. Remarkably this incommensurate spin-density wave (IC-SDW) order vanishes when the superconductivity dies at $H_{c2}$ \cite{Ken08,Kou10}, indicating that the magnetism and the exotic superconductivity are closely intertwined.  

The observation of the magnetic order in the HL phase calls for a reexamination of a simple FFLO interpretation.  To account for the coexistence of $d$-wave superconductivity and magnetic order confined exclusively in the superconducting state, several exotic superconducting orders, including pair-density wave state with a $\pi$-triplet component \cite{Agt09,Apr10,Miy07} and spatially inhomogeneous SDW state induced around the FFLO nodal planes \cite{Yan09},  have been proposed.  However, the nature of the HL phase is still unclear and under hot debate. 

In this study, to improve our understanding of how the exotic superconductivity and magnetism can interact in CeCoIn$_5$, we measured NMR spectra on the three distinct In sites in parallel fields.  The field evolution of the magnetization and the density of states (DOS) associated with the paramagnetic quasiparticle formation in the HL phase is extracted from the NMR spectra.  Our results provide strong evidence for the formation of the FFLO state via the second order phase transition, which coexists with the static magnetic order.

The $^{115}$In NMR measurements were performed for {\boldmath $H$}\,$\parallel$\,[100] in the field-cooling condition by using a phase-coherent spectrometer on high-quality single crystals ($T_c=2.3$\,K), whose physical properties are well characterized, as reported in the measurements of transport properties \cite{Nak}, specific heat, magnetic susceptibility \cite{Tayama}, thermal conductivity \cite{Kas05} and ultrasound velocities \cite{Wat04}. NMR spectra were obtained from a convolution of Fourier transform signals of the spin echo which was measured at 40$\sim$50\,kHz intervals. The Knight shift was obtained from central or satellite $^{115}$In lines ($I=9/2$) using a gyromagnetic ratio of 9.3295\,MHz/T and by taking into account parameters of the nuclear quadrupole interaction. 

The tetragonal crystal structure of CeCoIn$_5$ consists of alternating layers of CeIn$_3$ and CoIn$_2$ (Fig.\:\ref{fig:SDW}(b)). In a magnetic field {\boldmath $H$}\,$\parallel$\,[100], there are three inequivalent In sites.  The axially-symmetric In(1) is located in the CeIn layer, whereas In(2a) and In(2b) sites are located between Co and CeIn layers with the largest principal axis of electric field gradient parallel and perpendicular to the applied field, respectively (Fig.\:\ref{fig:SDW}(c)). The NMR frequency in the superconducting state is spatially-distributed and is determined by the local magnetic field and hyperfine coupling to the conduction electron spins: {\boldmath $H$}$_{\rm eff}$({\boldmath $r$})={\boldmath $H$}+{\boldmath $M$}$_O$({\boldmath $r$})+$A_{\rm hf}${\boldmath $M$}$_S$({\boldmath $r$}), where  {\boldmath $M$}$_O$({\boldmath $r$}) is the local magnetization due to the orbital (diamagnetic screening current) effect, $A_{\rm hf}$ is the hyperfine coupling constant \cite{Kuma2} and {\boldmath $M$}$_S$({\boldmath $r$})  is the local spin magnetization.  The Knight shift,  given as $K=A_{\rm hf}M_S/H$,  consists of the spin and orbital contributions, $K=K_{\rm spin} +K_{\rm orb}$.  The amplitude of $K_{\rm orb}$ is estimated at low field and low temperature in the superconducting state where $K_{\rm spin}$ vanishes. We note that in the vortex state of CeCoIn$_5$, line shapes at In(2a) and In(2b) sites with large amplitude of $A_{\rm hf}$  are predominantly determined by the hyperfine coupling ($A_{\rm hf}M_s \gg M_O$).

The antiferromagnetic staggered Ce moments due to IC-SDW with {\boldmath $\mu$}$_0 \parallel c$ induce  the in-plane hyperfine field perpendicular and parallel to the applied field {\boldmath $H$} at the In(2a) and In(2b) sites, respectively (Fig.\:\ref{fig:SDW}(c)), through dipolar type transferred hyperfine couplings \cite{Cur10}.  Consequently, the In(2b) spectra should be shifted (into two peaks) by the antiferromagnetic staggered field, whereas In(2a) spectra should not.  Figure\:\ref{fig:SDW}(d) depicts the field evolution of In(2b) spectra at 50\,mK as a function of Knight shift.   In the normal state for $\mu_0H\geq11.8$\,T, very sharp spectra  with line width less than 50\,kHz (black lines) are observed. For $\mu_0H<11.8$\,T the sharp In(2b) spectra broaden and split into two peaks with finite signal weight between them (red lines), which is characteristic of IC-SDW long-range order along one spatial dimension.  In the BCS phase below $H^*(\simeq 10$\,T), NMR spectra (blue lines) becomes a single asymmetric line as expected in the usual vortex state.  The inset of Fig.\:\ref{fig:SDW}(d) shows the field dependence of the internal field $H_{\rm int}$ determined by the difference of the resonance frequency of the two peaks in the HL phase. Above $H^*$,  $H_{\rm int}$ increases with $H$ and jumps to zero at the first-order $H_{c2}$ transition after reaching maximum value of $\sim0.16$\,T, which corresponds to $\sim0.15\,\mu_B$/Ce.  The vanishing of SDW order above $H_{c2}$ and the magnitude of $H_{\rm int}$ are consistent with the previous NMR \cite{You07,Cur10,Kou10} and neutron results \cite{Ken08,Ken10}.  

\begin{figure}[tb]
\includegraphics[width=0.78\linewidth]{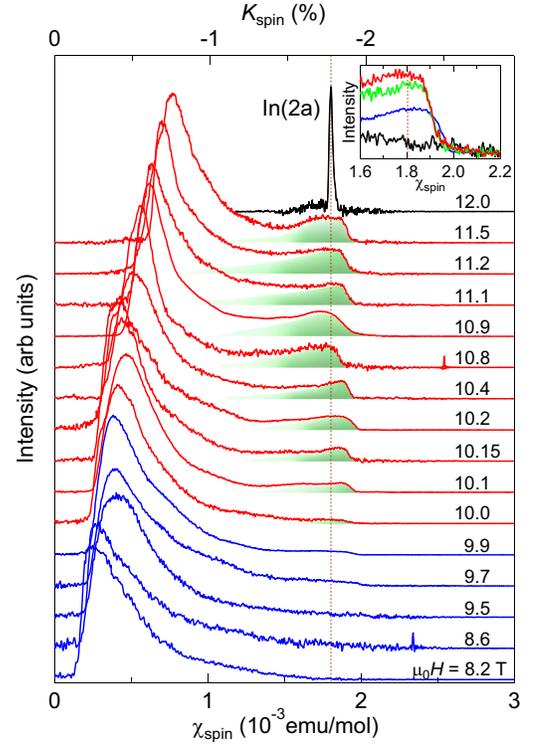}
\caption{
 (Color online) Field evolution of the NMR spectra at the In(2a) site at $T=0.05$\,K in the normal (black), HL (red), and BCS (blue) states. The integrated intensity of each spectrum below $H_{c2}$ is normalized.  The spin susceptibility (lower scale) is obtained by $\chi _{\rm spin}=K_{\rm spin}/A_{\rm hf}$ with $K_{\rm orb}=1.95\%$.  The green shaded region indicates the quasiparticle spectrum emerged in the HL phase. Inset: Blow-up of spectra near the edge structure at $\mu_0H=9.5$ (black), 10.2 (blue), 11.1 (green), and 11.2\,T (red). Dotted lines indicate the peak position in the normal state.
}
\label{fig:H_dep}
\end{figure}

The field evolution of the In(2a) spectra is depicted in Fig.\:\ref{fig:H_dep}. The sharp resonance line in the normal state (black) becomes antisymmetric broad lines  in the HL (red) and BCS (blue) phases.  The most salient feature of the HL-phase spectra is the emergence of the edge structure whose position coincides with the Knight shift in the normal state, as shown by the dotted line (also Fig.\:\ref{fig:H_dep}, inset).  This provides direct evidence for the emergence of the normal quasiparticle region in the HL phase \cite{Kak05,Ich07}.   We note that as the edge position is independent of magnetic field, the effect of the hyperfine field due to the SDW ordering is negligibly small. This indicates that In(2a) spectra are predominantly affected by the local spin susceptibility arising from the normal quasiparticles.

Here we comment on the reproducibility of the data.    The NMR spectra are reproducible for different rf powers and different cooling rates.  We also measured the spectra on several different single crystals.  Although the clear edge structure of  In(2a) spectra is observed in all crystals, its sharpness slightly depends on the crystal.  This appears to be related to the recent result that the HL phase is extremely sensitive to the nonmagnetic impurity \cite{Tok08}. The NMR spectra with less pronounced edge structure can also be found above $H^*$ in Ref.\:\onlinecite{Kou10}.  

\begin{figure}[t]
\includegraphics[width=0.78\linewidth]{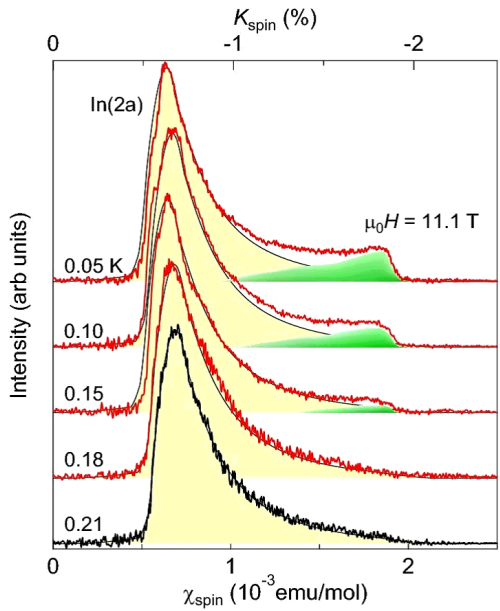}
\caption{
(Color online)
Temperature evolution of the In(2a) spectrum in HL (red) and BCS (black) phases at $\mu_0 H=11.1$\,T.  The peak intensity of each spectrum is normalized.  Thin black lines (yellow shaded region) indicate the spectrum at $T=0.21$\,K just above $T^*$.  The green shaded region in the low-temperature data indicates the quasiparticle spectrum formed in the HL phase. 
}
\label{fig:T_dep}
\end{figure}

We stress that the clear-cut sharp edge structure is consistent with the recent calculation of the NMR spectra in the FFLO state based on the microscopic Bogoliubov-de\,Gennes equation \cite{Ich07}.  For the quantitative analysis, we measured the temperature dependence of the In(2a) spectrum (Fig.\:\ref{fig:T_dep}).   As shown by thin black line, which is the spectrum at $T=0.21$\,K just above $T^*$, the shape of the main peak remains almost unchanged with temperature.  This allows us to separate the spectrum in the edge region (the green shaded region in Fig.\:\ref{fig:T_dep}) by subtracting the BCS spectrum at 0.21\,K (the yellow shaded region) from the whole spectra in the HL phase.  The green shaded regions in Fig.\:\ref{fig:H_dep} also indicate the spectra near the edge regions obtained by the similar method. 

\begin{figure}[tb]
\includegraphics[width=0.97\linewidth]{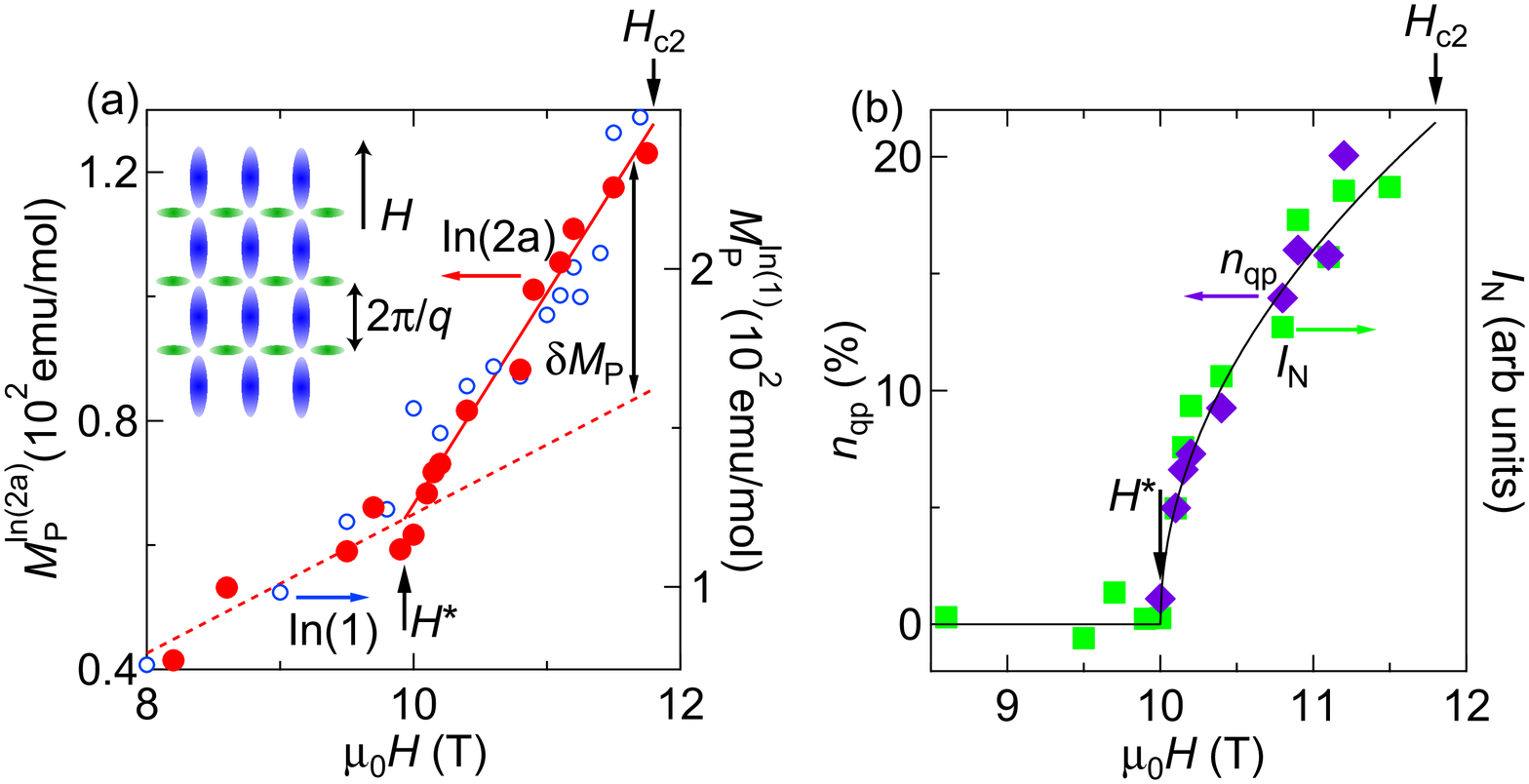}
\caption{
(Color online)
(a) Field dependence of the paramagnetic magnetization $M_p$ of the In(2a)  (red circles) and the In(1) (blue open circles) sites. Inset: Schematic quasiparticle structure in the FFLO state.  The nodal planes (green) with period of $2\pi/q$ appear perpendicular to the Abrikosov vortex lattice (blue).  (b) Field dependence of the DOS of the paramagnetic quasiparticles, $n_{qp}$ and $I_N$, extracted from the In(2a) spectra. The solid line is a fit to $\sqrt{H-H^*}$-dependence.
}
\label{fig:QP}
\end{figure}

The total paramagnetic magnetization $M_p$ is evaluated from the whole NMR spectrum $P(K_{\rm spin})$ by integrating the spin part of the Knight shift as $M_p(H)=\frac{H}{A_{\rm hf}}g(H)$, where $g(H)=\int K_{\rm spin}P(K_{\rm spin})dK_{\rm spin}/\int P(K_{\rm spin})dK_{\rm spin}$.  Figure\:\ref{fig:QP}(a) depicts the field dependence of $M_p$ obtained from the In(2a) spectra as well as $M_p(H)$ obtained from the same analysis for the In(1) spectra.   Both $M_p(H)$ from the In(1) and In(2a) spectra increase linearly with $H$ in the HL phase.   The difference of the paramagnetic magnetization between the HL and BCS phases, $\delta M_p(H)=M_p(H)-M_p^L(H)$,  increases continuously from zero as $\delta M_p(H) \propto (H-H^*)$ in the HL phase, where $M_p^L(H)$ is obtained by a linear extrapolation from the BCS phase (dotted line in Fig.\:\ref{fig:QP}(a)).  We emphasize that $M_p(H)$ cannot be obtained from the bulk magnetization measurements \cite{Tayama} which contains the contribution of the antiferromagnetic Ce moments.

The DOS of the normal quasiparticles emerged in the HL phase is proportional to the area of green shaded regions in Figs.\:\ref{fig:H_dep} and \ref{fig:T_dep}, as the intensity there should be proportional to the number of nuclei which detect the quasiparticle susceptibility.  Figure\:\ref{fig:QP}(b) depicts the field dependence of the normalized DOS, $n_{qp}$, which is obtained by the area of the green region divided by the area of the full spectrum at each field in Fig.\:\ref{fig:H_dep}.   To estimate the DOS in an alternative way, the intensity $I_N$ of the spectrum at the normal state Knight shift (dotted line in Fig.\:\ref{fig:H_dep})  is also plotted in Fig.\:\ref{fig:QP}(b).  Both $n_{qp}$ and $I_N$ increase in proportion to $\sqrt{H-H^*}$ in the HL phase. 

The $H$-linear dependence of the paramagnetic magnetization and $\sqrt{H}$-dependence of the low-energy DOS both provide key information about the order parameter that characterizes the emergence of the normal quasiparticles in the HL phase.  We stress that both field dependencies are exactly what are expected in the FFLO state where the order parameter is described by the amplitude of the modulation wave vector $q=\mid${\boldmath $q$}$\mid$.  The change in the paramagnetic magnetization, which is the first derivative of the free energy with respect to the applied magnetic field, is proportional to the square of the order parameter near the second order phase transition.  Then it increases in proportion to the magnetic field as $\delta M_p\propto q^2\propto (H-H^*)$ near the FFLO transition \cite{Ike072,Min07}.   Moreover, the DOS of the normal quasiparticles emerged in the FFLO phase is proportional to the number of nodal planes, which is proportional to $q$ (inset of Fig.\:\ref{fig:QP}(a)); $n_{qp}$ is expected to increase with $H$ as $n_{qp}\propto q \propto \sqrt{H-H^*}$.   Thus the quantitative analysis of the In(2a) and In(1) spectra  provide  strong support for the formation of the FFLO state.

It has been theoretically proposed that the IC-SDW moment is induced around the FFLO nodal planes \cite{Yan09}. In the presence of such a spatially non-uniform IC-SDW state, the NMR spectra at the In(2b) site in the HL phase consist of both contributions of the regions far outside (BCS spectrum with one peak) and around  the nodal planes (IC-SDW spectrum with two peaks).  In contrast, the present In(2b) spectra in the HL-phase shown in Fig.\:\ref{fig:SDW}(d) consist entirely of the split one with no discernible component of the BCS spectrum. This suggests rather uniform IC-SDW, in which the magnetic order is present even in the region far away from the nodal planes. This is consistent with the absence of satellite peaks in the neutron scattering experiments \cite{Ken10}.

We point out that the present results are incompatible with the pair-density wave scenario \cite{Agt09,Apr10}.  The emergence of the normal quasiparticles in the HL-phase is not expected in this scenario.  Moreover, the order parameter $q$, which depends on $H$ in accord with the FFLO second order transition, is at odds with this scenario that assumes the Cooper pair with the SDW modulation wave vector {\boldmath $Q$}$_S$, which is field independent \cite{Ken08}.

It has been reported that the magnetic moment induced by the SDW disappears when the magnetic field is tilted away from the $ab$ plane by 17$^\circ$ \cite{Bla10}.  Moreover, a possible FFLO phase with no magnetic order appear in {\boldmath $H$}\,$\parallel c$ \cite{Bia03,Kum06}.  These results appear to indicate that coupling between FFLO and IC-SDW becomes weaker with tilting {\boldmath $H$} from the $ab$ plane.  Recently, it has been suggested that in CeCoIn$_5$ with $d_{x^2-y^2}$-wave symmetry the strong Pauli paramagnetism plays an important role for the SDW formation as well as the FFLO state particularly in parallel field \cite{Ike10,Mac10}.  Further investigation is required to understand the perplexing relationship between the coexisting FFLO and IC-SDW states. 

In summary, $^{115}$In NMR measurements demonstrate the emergence of a spatially distributed normal quasiparticle region in the HL phase of CeCoIn$_5$ in parallel field.  The field evolution of the paramagnetic magnetization and low-energy quasiparticle DOS can be described well by the order parameter associated with the nodal plane formation via the FFLO second order phase transition.  The NMR spectra also reveal that the spatially uniform SDW coexists with the FFLO nodal planes. 

We thank N.~Curro, R.~Ikeda, K.~Ishida, K.~Machida, H.~Shimahara, and Y.~Yanase for valuable discussion.


\begin{thebibliography}{99}
\bibitem{Kuma2} The $A_{\rm hf}$ values of 1.8, -5.5 and -3.0\,$T/\mu_B$ for In(1), In(2a) and In(2b) site, respectively, are obtained from the $K$-$\chi$ plot below $T=2$\,K by using $K(T)$ in the normal state at $\mu_0H=12$\,T and $\chi(T)$ \cite{Tayama} for $H\parallel$\,[100].
\end{thebibliography}
\end{document}